\documentclass[twocolumn,showpacs,amssymb,aps,floatfix,draft]{revtex4}
\usepackage{epsf} 

\newcommand{\ketbra}[2]{\vert #1\rangle \! \langle #2\vert}
\newcommand{\proof}{\par\noindent\textit{Proof:\ }}
\newcommand{\QED}{\hfill$\square$\par\vskip24pt}
\newcommand{\tr}{\mathop\mathrm{Tr}\nolimits}
\newcommand{\F}{\mathbb{F}}
\newcommand{\flipt}{\widehat \F}
\newcommand{\identity}{\openone}

\newtheorem{The}{Theorem}

\newtheorem{Lem}[The]{Lemma}

\newcommand\beq{\begin{equation}}
\newcommand\eeq{\end{equation}}
\newcommand\bea{\begin{eqnarray}}
\newcommand\eea{\end{eqnarray}}
\newcommand\beas{\begin{eqnarray*}}
\newcommand\eeas{\end{eqnarray*}}

\begin{document}

\title{Asymptotic Relative Entropy of Entanglement for Orthogonally Invariant States}

\author{K. Audenaert}
\email{k.audenaert@ic.ac.uk}
\affiliation{QOLS, The Blackett Laboratory, Imperial College, London, SW7 2BW, UK}
\author{B. De Moor}
\affiliation{Dept. of Electrical Engineering (ESAT-SCD), KU Leuven, B-3010 Leuven-Heverlee, Belgium}
\author{K.G.\ H.\ Vollbrecht}
\email{k.vollbrecht@tu-bs.de}
\author{R.F.\ Werner}
\email{R.Werner@tu-bs.de}
\affiliation{Institut f\"ur Mathematische Physik, TU Braunschweig, 38106 Braunschweig, Germany}

\date{\today}

\begin{abstract}
For a special class of bipartite states we calculate explicitly
the asymptotic relative entropy of entanglement $E_R^\infty$ with
respect to states having a positive partial transpose (PPT). This
quantity is an upper bound to distillable entanglement. The states
considered are invariant under rotations of the form $O\otimes O$,
where $O$ is any orthogonal matrix. We show that in this case $E_R^\infty$
 is equal to
another upper bound on distillable entanglement, constructed by
Rains. To perform these calculations, we have introduced a number
of new results that are interesting in their own right: (i) the
Rains bound is convex and continuous; (ii) under some weak
assumption, the Rains bound is an upper bound to $E_R^\infty$;
(iii) for states for which the relative entropy of entanglement
$E_R$ is additive, the Rains bound is equal to $E_R$.
\end{abstract}


\maketitle
\section{Introduction}

In spite of the impressive recent progress in the theory of
entanglement \cite{entanglement}, many fundamental questions or challenges still remain
open. One of these issues is to decide whether a given state is
entangled or not. Another question is to find criteria for the distillability
of a state, i.e. whether pure state entanglement can be
recovered from the original state by means of local operations and classical
information exchange.

Since entangled states are a resource in many basic
protocols in quantum computation and quantum communication, a need has emerged to
quantify entanglement. This leads to more advanced
challenges: how much entanglement is needed to create a given
state and how much entanglement can be recovered?

Since these questions lead to very high dimensional optimization
problems, it is often helpful or even inevitable to restrict oneself to
states exhibiting a very high symmetry. The two most common
one-parameter families of symmetric states are the so-called
`Werner' states \cite{Werner89} and the `Isotropic' states, which are related to one another via the
partial transposition operation. A larger set of symmetric states,
containing these two sets as special cases, are the OO-invariant
states, which are the states considered in this paper.

So far it is not known how to calculate distillation rates for
arbitrary states, and even for symmetric states this optimization
seems to be intractable. One possible way to partially circumvent
this problem is to calculate good bounds for the distillation
rates. A well-known upper bound for the distillable entanglement
is the relative entropy of entanglement \cite{relent}, which is
itself defined as an optimization: $$ E_R(\rho)=\inf_{\sigma \in
{\cal D}} S(\rho||\sigma). $$ In this formula,
$S(\rho||\sigma)=\tr (\rho \log \rho-\rho\log \sigma)$ is the
relative entropy (the quantum mechanical analog of the
Kullback-Leibler divergence) and the minimum is taken over all
states $\sigma$ in the convex set $\cal D$. The relative entropy
between two states is a measure of distinguishability and can
intuitively be regarded as a kind of distance measure, although it
violates most of the axioms that are required of a distance
measure \cite{relent}. In the originally proposed definition of
the relative entropy of entanglement, $\cal D$ is the set of
separable states, so that the $E_R(\rho)$ expresses the minimal
distinguishability between the given state and all possible
separable states. When using the $E_R$ as an upper bound to
distillability, however, it is fruitful to enlarge the set $\cal
D$ to the set of states with positive partial transpose (the PPT
states) \cite{Rains_Lemma}. The corresponding minimal relative
entropy, {\em the relative entropy of entanglement with respect to
PPT states} (REEP), is generally smaller than the (separability)
relative entropy of entanglement while it still is an upper bound
to distillability; this is so because all PPT states have
distillability zero. Hence, the REEP is a sharper bound on the
distillability than the separability relent. This enlargement of
$\cal D$ has the additional benefit that the set of PPT states is
much easier to characterize than the set of separable states, for
which no general operational membership criterion exists.

Nevertheless, neither for the REEP nor for the relative entropy of
entanglement is there a general solution known of the optimization
problem for arbitrary states, not even for the otherwise simple
case of two qubits. However, the calculations become tractable
when restricting oneself to symmetric states.

Contrary to earlier conjectures, neither REEP nor the relative
entropy of entanglement is additive, i.e.\ $E_R(\rho_1 \otimes
\rho_2)\leq E_R(\rho_1)+E_R(\rho_2)$ is a strict inequality for
some states. It is expected, however, that this non-additivity
becomes less severe for the {\em asymptotic relative entropy of
entanglement with respect to PPT states}(AREEP), which is defined as the regularisation
$$
E_R^\infty(\rho)=\lim_{n\rightarrow\infty}\frac{1}{n}
E_R(\rho^{\otimes n }),
$$
and which at the same time provides yet a sharper bound to
distillable entanglement.

The calculation of the AREEP has first been done on Werner states
\cite{ka}, showing that the asymptotic value can be a good deal
smaller than the single-copy value. Surprisingly, it turns out
that on Werner states the AREEP is equal to another upper bound on
distillability, the so-called {\em Rains bound}
\cite{Rains_semidef}
\beq
\label{REntropy}
R(\rho)=\inf_{\sigma}S(\rho||\sigma)+\log\tr|\sigma^{T_2}|.
\eeq
One of the things we will show in this paper
is that this equality remains valid over the larger class of
OO-invariant states.

To calculate the AREEP on OO-invariant states in a relatively
simple way, we will make use of four ingredients:
\begin{itemize}
\item
First of all, REEP is additive on a large part of the state space.
This will be discussed in Sec.\ \ref{sec_add_relent}. For this
additive region, the calculation of the AREEP is trivial, as the
(single-copy) REEP for OO-invariant states has been calculated
before.

\item We will make use of the convexity of the AREEP (recollected in Sec.\ \ref{sec_convex_relent})
and of the Rains bound (proven in Sec.\ \ref{sec_convex_rains}).
In Sec.\ \ref{sec_convex_relent} we use this convexity to define
the ``minimal convex extension'' of the AREEP from the additive
areas to the full state space.

\item In Sec.\ \ref{sec_rains_relent} we will present a close connection between
the Rains Bound $R(\rho)$ and the AREEP. We will establish an
upper bound to the AREEP that will turn out to be tight on
OO-invariant states.

\item In Sec.\ \ref{sec_OO_sym} we will recall the basic properties of
OO-invariant states resulting from their symmetry. It is exactly
this symmetry that makes the calculation feasible.
\end{itemize}
Using these results, we will give a complete calculation of the
AREEP of OO-invariant states in Sec.\ \ref{Sec_OO} and prove that
this quantity is equal to the Rains bound for these states. We
will summarise the results of the paper in Sec.\ \ref{sec_disc}
and state a number of open problems.
\section{Additivity of relative entropy of entanglement} \label{sec_add_relent}
The additivity of the REEP was a folk conjecture, supported by
various numerical calculations and analytical case studies.
Nevertheless, it turned out to be wrong \cite{vollbrecht1}. The
misleading numerical result can be explained in hindsight by the
fact that, indeed, in great parts of the state space the REEP is
perfectly additive; the non-additive regions seem to be negligible
in size compared to the whole state space.

The following Lemma of Rains \cite{Rains_Lemma} can be utilized to
pinpoint regions where the REEP is additive.
\begin{Lem}[Rains-Additivity]\label{LemRains}
Let $\rho$ be a state and $\sigma$ a PPT state, such that
$E_R(\rho)=S(\rho||\sigma)$ and $[\rho,\sigma]=0$.
 If the condition
\beq
\label{cond3}
|(\rho \sigma^{-1})^{T_2}|\leq \identity
\eeq
holds, then the REEP is weakly additive on $\rho$, i.e.,
$E^\infty_R(\rho)=E_R(\rho)$. If it satisfies the stronger
condition
\beq
\label{cond33}
0 \leq(\rho \sigma^{-1})^{T_2}\leq\identity
\eeq
then REEP is strongly additive, i.e.,
$E_R(\rho \otimes \tau)=E_R(\rho)+E_R(\tau)$ holds for an
arbitrary state $\tau$.
\end{Lem}

Knowing the optimal $\sigma$ for a given state $\rho$, it is
straightforward to check condition (\ref{cond3}). Checking the
additivity therefore only requires one to calculate the REEP.
\section{Convexity of the asymptotic relent}\label{sec_convex_relent}
By definition, the asymptotic version of a given quantity inherits
most of the important properties directly from its single-copy
``parent'' quantity. One such property, which will turn out to be
very helpful to calculate the AREEP, is convexity. The REEP itself
is known to be convex, but it is not obvious that quantities of
the form $E_n(\rho):=E(\rho^{\otimes n})/n$ should be convex
functions in $\rho$ too and, in fact, this does not hold in
general. Although convexity might not hold for finite $n$, for the
REEP it becomes valid again in the asymptotic limit.

\begin{Lem} \cite{rudolph}
Let $E$ be a positive, subadditive, convex and tensor-commutative functional
on the density matrices of a Hilbert space. Then the asymptotic measure
$E^\infty(\rho):=\lim_{n\rightarrow
\infty}\frac{1}{n}E(\rho^{\otimes n})$ exists and is convex and
subadditive.
\end{Lem}

In the first calculation of the AREEP \cite{ka} great effort was
necessary to construct a lower bound to AREEP. Utilizing the
convexity we are now able to do this in a much simpler way.
Indeed, for any convex (differentiable) function $f$, a lower
bound to $f$ is given by any of its tangent planes
$$
f(x)\geq f(y)+\nabla f(y) (x-y).
$$
Given an open subset $D$ where the function $f$ is known, we
can define the ``minimal convex extension'' of the function by
$$
\bar f(x)=\sup_{y \in D} f(y)+\nabla f(y) (x-y).
$$
Note that $\bar f$ is equal to $f$ on $D$. Furthermore, $\bar f$ is smaller or equal
than any convex function that equals $f$ on $D$. As a maximum
over affine functions it is itself convex.

To make this bound a good candidate for an estimation to the
AREEP, we need to know the AREEP on a sufficiently large part of
the state space. In fact the AREEP is easy to calculate on PPT
states, where it is simply zero. But this is obviously too trivial
a result, because this gives a lower bound equal to zero on the
whole state space. The next greater set for which we can easily
calculate the AREEP is the set of states where $E_R^\infty$ is
additive. A subset of this set can be found using the Lemma of
Rains. It will turn out that this subset is large enough to yield
a bound that equals $E_R^\infty$ (at least for OO-invariant
states).
\section{Convexity and continuity of the Rains bound}\label{sec_convex_rains}
Although the function that is to be minimized in Rains' bound, $S(\rho||\sigma)+\log\tr|\sigma^{T_2}|$,
is not convex in $\sigma$ over state space,
the minimum itself turns out to be convex in $\rho$.
We prove this by first showing that the minimization problem in the calculation of the Rains bound
can be converted to a {\em convex} problem.

To begin with, we can add a third term to the function to be
minimized, namely $-\log\tr[\sigma]$, because this term is zero
anyway. Secondly, we can enlarge the set over which one has to
minimize from the set of normalized states to the set ${\cal S} = \{s\ge 0,
\tr[s]\le 1\}$. This is so because the sum of the first two terms
is independent of $\tr[\sigma]$ and the third one monotonously
decreases with $\tr[\sigma]$; hence, the minimal value must be
found on the boundary of $\cal S$ corresponding with
$\tr[\sigma]=1$ and is, therefore, equal to the original minimum.
The second and third term can now be absorbed in
the first term: $S(\rho||\sigma)+\log\tr|\sigma^{T_2}|-\log\tr\sigma
= S(\rho||\sigma(\tr\sigma/\tr|\sigma^{T_2}|))$. Defining
$$
\tau = \sigma(\tr\sigma/\tr|\sigma^{T_2}|),
$$
it is easy to check that
$\sigma\in{\cal S}$ if and only if $\tau\in{\cal T} = \{t\ge 0,
\tr|t^{T_2}|\le 1\}$. Hence, the calculation of the Rains bound
has been transformed to the minimization problem
$$
R(\rho) = \min_{\tau\in{\cal T}} S(\rho||\tau).
$$
The importance of this
transformation stems from the fact that the resulting optimization problem is
a so-called convex optimization problem: the function to be
minimized is now convex in $\tau$, while the set over which the
minimization is performed is still convex. The latter statement
follows directly from the convexity of the negativity. Indeed, if
$\tau_1$ and $\tau_2$ are in $\cal T$, then they are positive and
have negativity $\le 1$. Hence, any convex combination of $\tau_1$
and $\tau_2$ is positive and has negativity $\le 1$ as well, and,
therefore, belongs to the set $\cal T$.

It is now easy to prove continuity and convexity of the Rains bound itself.
Continuity follows by noting that the proof of continuity of the quantity
$\inf_{\sigma\in\cal D} S(\rho||\sigma)$ in \cite{continuity}, where $\cal D$
is a compact convex set of {\em normalized} states containing the maximally mixed state,
does actually not depend on the trace of the various $\sigma$ in $\cal D$.
Hence, the theorem is also true for convex sets $\cal D$ containing non-normalized states,
and, specifically, for the set $\cal T$.

Convexity is also proven in the standard way, as has been done for $E_R$ \cite{relent}.
The standard proof again depends only on the convexity of the feasible set
and not on the normalization of the states it contains.

In this way we have proven:
\begin{Lem}
The calculation of the Rains bound can be reformulated as a convex minimization problem:
$$
R(\rho) = \min\{S(\rho||\tau): \tau\ge 0, \tr|\tau^{T_2}|\le 1\}.
$$
The Rains bound itself is a continuous and convex function of $\rho$.
\end{Lem}

\section{Relation between Rains' bound and the AREEP}\label{sec_rains_relent}
The results of the calculation of the AREEP on Werner states
suggests \cite{rains_private} that this quantity might be
connected with the quantity (\ref{REntropy}) defined by Rains,
and, moreover, that there are connections between the minimizing
$\sigma$ in Rains' formula and the asymptotic PPT state $\sigma$
appearing in $E_R^\infty$. Indeed, it turns out that one can give
a simple relation between these two quantities, if we require as
an additional restriction that $\sigma$ in (\ref{REntropy})
satisfies $|\sigma^{T_2}|^{T_2}\geq 0$. If the restriction does
not hold the Lemma might still be true, but we have not been able
to prove this.

\begin{Lem}\label{RP}
An upper bound for the AREEP is given by \bea
R'(\rho)&:=&\inf^*_{\sigma}S(\rho||\sigma)+\log(\tr|\sigma^{T_2}|)\nonumber \\
\label{rns}
&\geq& E^\infty_R (\rho),
\eea
where the asterisk means that the infimum is to be taken over all states $\sigma$ satisfying
\beq
\label{cond1}
|\sigma^{T_2}|^{T_2}\geq 0.
\eeq
\end{Lem}
We will refer to the quantity $R'(\rho)$ as the {\em modified Rains bound}.
The set of states satisfying condition (\ref{cond1}) is easily
seen to be convex, so that the modified Rains bound is also
continuous and convex.
\proof
It can easily be seen that the Lemma is valid if we restrict
$\sigma$ to be a PPT state, since then the second term in (\ref{rns}) vanishes and
we get the trivial inequality $E^\infty_R(\rho)\leq E_R(\rho)$.
This means that we can restrict ourselves to the case where $\sigma$ is a
non-PPT state, i.e.\ $\tr|\sigma^{T_2}|>1$.

Let $\sigma$ be an arbitrary non-PPT state such
that $\bar\sigma:=|\sigma^{T_2}|^{T_2}\geq 0$, then
$$
\sigma_n=\frac{\sigma^{\otimes n}+\bar \sigma^{\otimes n}}{1+(\tr{\bar \sigma})^n}
$$
is a PPT-state. Taking this PPT state as a trial state in the
optimization for the AREEP, we get \bea E_R(\rho^{\otimes
n})&\leq& S(\rho^{\otimes n}||\sigma_n)= S\left(\rho^{\otimes
n}||\frac{\sigma^{\otimes n}+
\bar\sigma^{\otimes n}}{1+(\tr{\bar \sigma})^n}\right) \nonumber \\
&\leq& S\left(\rho^{\otimes n}||\frac{\sigma^{\otimes n}}{1+(\tr{\bar \sigma})^n}\right) \label{l3}\\
&=& n S(\rho||\sigma)+\log(1+(\tr\bar\sigma)^n). \nonumber
\eea
In (\ref{l3}) we have used the fact that the relative entropy is
operator anti-monotone in its second argument (Corollary 5.12 of
\cite{Ohya}), i.e.\ $S(\rho||\sigma+\tau)\leq S(\rho||\sigma)$ for positive $\tau$.
Taking the limit $n\rightarrow\infty$ and using $\tr\bar\sigma>1$ we get
\bea
E^\infty_R(\rho)&=&\lim_{n\rightarrow\infty}
\frac{1}{n}E_R(\rho^{\otimes n}) \nonumber \\
&\leq& \lim_{n\rightarrow\infty}
 S(\rho||\sigma)+\frac{\log(1+(\tr\bar\sigma)^n)}{n} \nonumber \\
 &=&S(\rho||\sigma)+\log\tr\bar\sigma. \label{l18}
 \eea
In order to get the best bound, we take the minimum over all feasible states
$\sigma$ in equation (\ref{l18}), giving
$$
E^\infty_R(\rho)\leq \inf^*_\sigma S(\rho||\sigma)+\log\tr|\sigma|^{T_2}
$$
where the infimum is taken over all states $\sigma$ satisfying $|\sigma^{T_2}|^{T_2}\geq 0$.
\QED

It is easy to see that for PPT states $\sigma$, $|\sigma^{T_2}|^{T_2}\geq
0$. Hence, the feasible set in the minimization of $E_R$ is a subset of the one for $R'$, which is
again a subset of the one for $R$. Therefore, we have the
inequalities
$$
R(\rho)\leq R'(\rho)\leq E_R(\rho).
$$

We also have the following Theorem:
\begin{The}
For $E_R$-additive states $\rho$ (i.e.\
$E_R(\rho)=E_R^\infty(\rho)$), the Rains bound is equal to the
AREEP and is additive.
\end{The}
\proof We have, in general, $R'(\rho)\leq E_R(\rho)$. On the other
hand, for additive states $E_R(\rho)=E_R^\infty(\rho)$, and
$E_R^\infty(\rho)\leq R'(\rho)$ by the Lemma 4. Therefore,
$R'(\rho)=E_R(\rho)=E_R^\infty(\rho)$ for all additive $\rho$.
This also implies that the PPT state $\sigma$ that is optimal for
$E_R$ is also optimal for $R'$.

To show that $R$ is also equal to $E_R$, we need to show that this
$\sigma$ is optimal for $R$ as well.
We use the reformulation of the Rains bound as a convex
minimization problem $R(\rho) = \min_\tau\{S(\rho||\tau): \tr|\tau^{T_2}|\le 1 \}$.
For the modified Rains bound, we have the additional restriction on
the feasible set that $|\tau^{T_2}|^{T_2}\ge 0$.
For clarity, let us write $\tau$ for the optimal $\tau$ for $R$ and $\tau'$ for the
optimal one for $R'$.
We have to show that $\tau=\tau'$, i.e.\ that $\tau$ is in the set
for which $|\tau^{T_2}|^{T_2}\ge 0$.

Suppose $\tau$ were
outside this set, then, following a general property of convex optimization problems,
$\tau'$ would have to be on the boundary of the set,
i.e.\ $|\tau'^{T_2}|^{T_2}$ would have to be positive and
rank-deficient.
On the other hand, we already showed that the optimal $\sigma'$ for $R'$ for
additive $\rho$ must be PPT, so that $\tau'=\sigma'$ and $|\sigma'^{T_2}|^{T_2} =
\sigma'$. Therefore, the rank-deficiency of $|\tau'^{T_2}|^{T_2}$ implies that $\sigma'$ itself
should be rank-deficient. However, if $\rho$ is not itself rank-deficient, then this cannot be,
because $\sigma'$ appears as second argument in the relative entropy and would then
give an infinite relative entropy, contrary to the statement that
$\sigma'$ actually minimises it. This proves that $R'(\rho)=R(\rho)$ for full-rank,
additive $\rho$. By continuity of the Rains bound this must then also hold for rank-deficient $\rho$.

Additivity of $R$ for $E_R$-additive states follows by regularising both sides of the equality
$R(\rho)=E_R^\infty(\rho)$, and noting that the right-hand side does not change.
\QED

We have introduced the operation $\sigma\mapsto|\sigma^{T_2}|^{T_2}$ as
a mathematical tool, and we doubt whether it has any
real physical significance (as was the case for the partial transpose). Nevertheless, its usefulness is
apparent from the above Lemma. A natural question to ask is
whether there really are states $\sigma$ for which $|\sigma^{T_2}|^{T_2}$
is not positive. We call states like this {\em binegative} states.
If they would not exist, then the modified Rains bound would just
be equal to the original Rains bound.
We have performed numerical investigations that have shown that, indeed,
binegative states exist, provided the dimensions of the system are higher than $2\times 2$.
For $2\times 2$ systems, extensive calculations failed to produce binegative states, which
suggests they might not exist in such systems.
For higher dimensions, binegative states have been produced, and they
always appear to be located close to the boundary of state space, i.e.\
have a smallest eigenvalue which is very small. In the present
setting, this is good news, because it implies that the modified
Rains bound will typically be close to the original Rains bound.

As one of the few exact results on the existence of binegative states,
we have been able to prove that pure states are never binegative:
\begin{Lem}
For any pure state $\psi$, $\big|\ketbra{\psi}{\psi}^{T_2}\big|^{T_2} \ge 0$.
\end{Lem}
\proof
Let $\psi$ have a Schmidt decomposition $\psi = \sum_i \lambda_i
u_i \otimes v_i$, then $\ketbra{\psi}{\psi}^{T_2} = \sum_{i,j}
\lambda_i \lambda_j \ketbra{u_i}{u_j}\otimes\ketbra{v_i}{v_j}^T$
and, exploiting the orthogonality of the vectors $u_i$ and of the
vectors $v_j$,
\beas
\big|\ketbra{\psi}{\psi}^{T_2}\big| &=&
\big(\sum_{i,j,k,l} \lambda_i\lambda_j\lambda_k\lambda_l
\ketbra{u_i}{u_j}\ketbra{u_k}{u_l} \otimes \\
&& (\ketbra{v_k}{v_l}\ketbra{v_i}{v_j})^T \big)^{1/2} \\
&=&
\big(\sum_{i,j} (\lambda_i\lambda_j)^2
\ketbra{u_i}{u_i} \otimes
\ketbra{v_j}{v_j}^T\big)^{1/2},
\eeas
since only the terms with $i=l$ and $j=k$ survive.
Again by orthogonality, taking the square root amounts to removing
the square on the factor $(\lambda_i\lambda_j)^2$.
Now, one clearly sees that the resulting expression corresponds to a product
state; hence, the partial transpose is still a state, which proves that
$\psi$ is not binegative.
\QED

In Section \ref{sec_OO_sym} we will show that no OO-invariant
state is binegative either. We will see that condition (\ref{cond1}) will be
fulfilled for the states we are considering in this paper.
Therefore, we will henceforth make no distinction anymore between $R$ and $R'$.
\section{OO-Invariant states}\label{Sec_OO}
We will now apply the tools obtained in the previous sections to
the complete calculation of the AREEP of OO-invariant states.
\subsection{Calculating the AREP on Werner states}
To illustrate how the calculation of the AREEP on OO-invariant
states will proceed, we apply the method first on Werner states,
reproducing the results of \cite{ka}.

Werner states can be written as
$$
\rho(p)=p \frac{P_-}{r_-}+(1-p)\frac{P_+}{r_+},
$$
where $P_+(P_-)$ denotes the normalized projection onto the symmetric
(antisymmetric) subspace of dimension $r_{\pm}=\frac{d^2\pm d}{2}$
and $p$ is a real parameter ranging from $0$ to $1$.

First of all, we need to know $E_R$ on these states. All states
with $p\leq \frac{1}{2}$ are PPT and, therefore, have both $E_R$
and $E_R^\infty$ equal to zero. For all non-PPT Werner states
$p>\frac{1}{2}$, the minimizing PPT-state is the state with
$p=\frac{1}{2}$. Knowing this state, we can easily write down the
REEP for all Werner states. To calculate the AREEP we use the
three steps introduced in the previous three Sections.

In the first step we use the lemma of Rains and check the
additivity condition (\ref{cond3}). An easy and straightforward
calculation leads to the result that all Werner states
satisfying $p\leq \frac{1}{2}+\frac{1}{d}$ are
additive and, therefore, have $E_R$ equal to $E_R^\infty$.

In the second step we calculate the Rains bound for Werner states.
Due to the high symmetry this is an easy task, already done by
Rains \cite{Rains_semidef}. In fact, we do not need to compute the Rains bound for all
states. For our purposes, we will only need the
Rains bound for $p=1$.

In the last step we calculate the tangent to the REEP in the point
$p=\frac{1}{2}+\frac{1}{d}$, which gives us the minimal convex
extension for all states with $p>\frac{1}{2}+\frac{1}{d}$. It
turns out that this minimal extension touches the Rains bound
again in the point $p=1$. This is sufficient to prove that the
minimal convex extension is equal to $E_R^\infty$ everywhere.
Indeed, by the convexity of $E^\infty_R$ the tangent yields a
lower bound and, furthermore, also implies that the tangent is an
upper bound between $p=1$ and $p=\frac{1}{2}+\frac{1}{d}$, because
at the end-points it equals $E^\infty_R$.

In fact, for Werner states, the same result can easily be obtained by
the observation that the Rains bound and the minimal convex
extension are equal on the whole range of $p$. But for
OO-invariant states the task to prove equality of these two quantities will become quite difficult.
Fortunately, we can
restrict ourselves to prove equality only on the border of the
state space as this will be sufficient for the calculation.
Equality of the Rains bound and
$E_R^\infty$ on the whole state space will follow automatically
from the convexity of both quantities.

We will now turn to the calculation for the OO-invariant states.
\subsection{Using symmetries} \label{sec_OO_sym}
The class of states we want to look at
commute with all unitaries of the form $O \otimes O$, where $O$ is
an orthogonal matrix. These so-called OO-invariant states lie in the
commutant $G'$ of the group $G=\{O\otimes O\}$. The commutant is
spanned by three operators, the identity operator $\identity$, the Flip
operator $\F$ defined as the unique operator for which $\F \psi \otimes \phi=\phi \otimes \psi$
for all vectors $\psi$ and $\phi$,
and the unnormalized projection on the maximally entangled state
$\flipt=\sum_{ij}\ketbra{ii}{jj} = d\ketbra{\Psi}{\Psi}$; here, $d$ is the dimension of either subsystem.
Every operator contained in this commutant can be written as a linear
combination of these three operators. To be a proper state such an
operator has to fulfill the two additional constraints of
positivity and normalization.

As coordinates parameterizing the OO-invariant states, we choose the
expectation values of the three operators $\identity,
\F$ and $\flipt$ in the given state. The expectation value of the identity, $\big< \identity\big>_\rho$, gives us
just the normalization, so we are left with the two free parameters
$f:=\big<\F\big>_\rho$ and $\hat f:=\big<\flipt\big>_\rho$.
For future reference, we collect the basic formulae here for performing calculations in this
representation.

The traces of the basis operators are given by
\beas
\tr[\identity] &=& d^2 \\
\tr[\F] &=& d \\
\tr[\flipt] &=& d.
\eeas
The inner products between them are easily calculated from the relations
\beas
\F^2 &=& \identity \\
\F\flipt &=& \flipt\F = \flipt \\
\flipt^2 &=& d\F.
\eeas
From this basis $\{\identity,\F,\flipt\}$, an orthogonal basis of projectors
can be constructed. The operator $\F$ is not positive and can be written as $\F=\F_+ - \F_-$; here
$\F_+$ and $\F_-$ denote the positive and negative part of $\F$, respectively, and are defined by
the equations $X=X_+ - X_-$, $|X|=X_+ + X_-$ (note that both the positive and negative part are
positive by this definition).
Since $\F^2=\identity$, $\F_+ + \F_- = \identity$, and $\F_- = (\identity-\F)/2$.
Furthermore, as $\F\flipt = \flipt$, $\flipt<\F_+$. Therefore,
the following operators form an orthogonal set of projectors and add up to the identity:
\beas
U&=& \flipt/d \\
V&=& (\identity-\F)/2 \\
W&=& (\identity+\F)/2 - \flipt/d.
\eeas
The traces of these projectors are
\beas
\tr [U]&=&1 \\
\tr [V]&=&d(d-1)/2 \\
\tr [W]&=&(d+2)(d-1)/2.
\eeas
The original basis is
related to the orthogonal one by
\beas
\identity &=& U+V+W \\
\F &=& U-V+W \\
\flipt &=& d U.
\eeas

For a general OO-invariant $\rho$, we write
$$
\rho = a\identity +b\F+c\flipt.
$$
The relation between the coefficients $a$, $b$, $c$ and $f$ and $\hat f$ is given by
$$
\left[\begin{array}{c}1\\f\\\hat f\end{array}\right] =
d\left[\begin{array}{ccc}d&1&1\\1&d&1\\1&1&d\end{array}\right]
\left[\begin{array}{c}a\\b\\c\end{array}\right],
$$
and, inversely, by
$$
\left[\begin{array}{c}a\\b\\c\end{array}\right] =
\frac{1}{d(d-1)(d+2)}\left[\begin{array}{ccc}d+1&-1&-1\\-1&d+1&-1\\-1&-1&d+1\end{array}\right]
\left[\begin{array}{c}1\\f\\\hat f\end{array}\right].
$$
In terms of the orthonormal basis, $\rho$ can be written as
\beq
\label{eq:UVW}
\rho = \frac{\hat f}{d}U+\frac{1-f}{d(d-1)}V+\frac{d+df-2\hat f}{d(d-1)(d+2)}W.
\eeq
Positivity of $\rho$ thus amounts to the conditions
\beas
0 &\le& \hat f\\
f &\le&1 \\
\hat f &\le& d(1+f)/2.
\eeas
The representation of the partial transpose of $\rho$ is very easy, since $\F$ and $\flipt$ are
just each other's partial transpose. Hence, the partial transpose of $\rho$ is obtained by swapping
$\F$ and $\flipt$. In the basis $\{\identity,\F,\flipt\}$, taking the partial transpose corresponds, therefore,
to interchanging the parameters $f$ and $\hat f$.
The partial transposes of the projectors $U$, $V$ and $W$ are
easily calculated to be
\beas
U^{T_2} &=& \frac{1}{d}(U-V+W) \\
V^{T_2} &=& \frac{1-d}{2}U + \frac{1}{2}V + \frac{1}{2}W \\
W^{T_2} &=& \big(\frac{1+d}{2}-\frac{1}{d}\big) U
+\big(\frac{1}{2}+\frac{1}{d}\big) V
+\big(\frac{1}{2}-\frac{1}{d}\big) W.
\eeas

\begin{figure}
\begin{center}
\epsfxsize=7cm
\epsffile{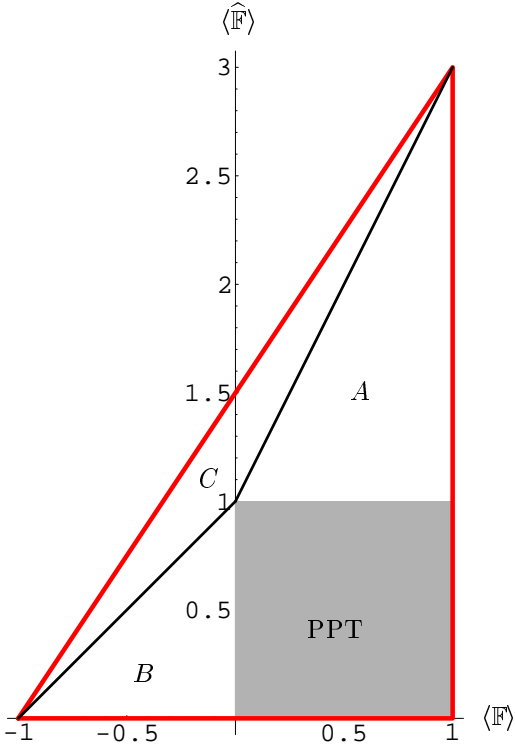} \caption{ State space of OO-invariant states
(case $d=3$). These states are parameterized by the two parameters
$f=\langle\F\rangle$ and $\hat f =\langle \flipt\rangle$. The
outer triangle represents the values corresponding to states
(positivity). The grey area is the set of PPT OO-invariant states.
The region of non-PPT states is subdivided further in the three
triangular regions labeled $A$, $B$ and $C$. For each of these
regions the optimal $\sigma$ appearing in the definition of the
REEP is of a different form. } \label{fig1}
\end{center}
\end{figure}

From these formulae one can see that the set of OO-invariant states constitutes a triangle in the
$(f,\hat f)$ parameter space, as plotted in Figure \ref{fig1}. Taking the partial transpose amounts to
taking the mirror image around the line $f=\hat f$. Therefore, the set of PPT states are those contained in the
grey square $0\le f,\hat f\le 1$ in Figure \ref{fig1}.

What will make the calculation of the REEP easy for these
OO-invariant states is the existence of a `twirl' operation
\cite{Werner89}, a projection operation $T$ that maps an arbitrary
state $\rho$ to an OO-invariant state $T(\rho)$ and that preserves
PPT-ness, i.e., that maps every PPT state to an OO-invariant PPT
state. Since
$$
S(\rho||\sigma)\geq S(T(\rho)||T(\sigma))
$$
this guarantees that the minimum relative entropy for an
OO-invariant state is attained on another OO-invariant PPT state
\cite{Rains_Lemma,vollbrecht1}. Hence, we can reduce the very
high-dimensional optimization problem to an optimization in our
two-dimensional OO-invariant state space. This optimization has
been done \cite{vollbrecht1} and the minimizing PPT states are as
follows. Let a state $\rho$ be determined by the expectation
values $\big< \F \big>_\rho=f$ and $\big<\flipt \big>_\rho=\hat
f$. Similarly, let the expectation values in the optimizing PPT
state $\sigma$ be given by $\big< \F \big>_\sigma=s$ and
$\big<\flipt \big>_\sigma=\hat s$. Then the following table gives
the expressions for $s$ and $\hat s$, depending on which region
the state $\rho$ is in:
\begin{center}
\begin{tabular}{ccc}
  \hline
  Region & $s$ & $\hat s$ \\ \hline
  $A$ & $\frac{1+(d-1)f-\hat f}{d-\hat f}$ & $1$ \\
  $B$ & $0$ & $\frac{\hat f}{1+f}$ \\
  $C$ & $0$ & $1$ \\
  \hline
\end{tabular}
\end{center}

To end this section, we give the formulas for the relative entropy
 and the negativity of OO-invariant states. Let the
states $\rho$ and $\sigma$ be determined by their expectation
values $f, \hat f$ and $s, \hat s$, respectively. Using the state
representation (\ref{eq:UVW}), in the orthogonal basis
$\{U,V,W\}$, the relative entropy of $\rho$ w.r.t.\ $\sigma$ is
given by \bea S(\rho||\sigma) & = & \frac{\hat f}{d}
\log\Big(\frac{\hat f}{\hat s}\Big)\tr U +
\frac{1-f}{d(d-1)} \log\Big(\frac{1-f}{1-s}\Big) \tr V \nonumber \\
&& +\frac{d+df-2\hat f}{d(d-1)(d+2)} \log\Big(\frac{d+df-2\hat f}{d+ds-2\hat s}\Big) \tr W \nonumber \\
&=&
\frac{\hat f}{d} \log\frac{\hat f}{\hat s} +
\frac{1-f}{2} \log\frac{1-f}{1-s} \nonumber \\
&& +\frac{d+df-2\hat f}{2d} \log\frac{d+df-2\hat f}{d+ds-2\hat s}.
\label{relent1}
\eea
Recollecting that taking the partial transpose corresponds to interchanging $s$ and $\hat s$,
the negativity of $\sigma$ is given by
\bea
\tr |\sigma^{T_2}| &=&
\left|\frac{s}{d}\right| \tr U +
\left|\frac{1-\hat s}{d(d-1)}\right| \tr V \nonumber \\
&& + \left|\frac{d+d\hat s-2s}{d(d-1)(d+2)}\right| \tr W \nonumber \\
&=& \frac{|s|}{d} + \frac{|1-\hat s|}{2} + \frac{|d+d\hat
s-2s|}{2d}. \label{negativity}
\eea
The positivity condition on $\sigma$ implies that the absolute value sign on the third term is superfluous.

In a similar way, we can show that for any OO-invariant state
$\sigma$, the operator $|\sigma^{T_2}|^{T_2}$ is a state again, as we had promised.
Indeed,
\beas
|\sigma^{T_2}|^{T_2} &=& \left|\frac{s}{d}\right| U^{T_2} +
\left|\frac{1-\hat s}{d(d-1)}\right| V^{T_2} \\
&& + \left|\frac{d+d\hat s-2s}{d(d-1)(d+2)}\right| W^{T_2}.
\eeas
An easy but somewhat lengthy calculation shows that this expression can be rewritten in terms of
$U$, $V$ and $W$ with positive coefficients.
\subsection{Additive Areas}
In the first step we want to identify the areas within the state
space triangle where the REEP is additive.
\begin{Lem}
$E_R(\rho)$ is additive for all OO-states satisfying
$\big< \F \big>\geq \frac{-2}{d}$ and
$\big< \flipt \big> \leq 3-\frac{4}{d} + (d-1)\big< \F \big> $.
\end{Lem}
\proof
Utilizing Lemma \ref{LemRains} we only have to check condition
(\ref{cond3}) for every OO-invariant state $\rho$ and the corresponding
optimal PPT-states $\sigma$.
In the $\{U,V,W\}$-basis, $\rho\sigma^{-1}$ is directly given by
$$
\rho\sigma^{-1} = uU + vV + wW,
$$
with
\beas
u &=& \frac{\hat f}{\hat s} \\
v &=& \frac{1-f}{1-s} \\
w &=& \frac{d+df-2\hat f}{d+ds-2\hat s}.
\eeas
In order to perform the partial transpose, we replace $U,V,W$ by their partial transposes and express
them in the original $U,V,W$ again.
This yields
$$
(\rho\sigma^{-1})^{T_2} = (a+bd+c)U + (a-c)V+(a+c)W,
$$
with
$$
a = \frac{w+v}{2}, \,\,
b = \frac{w-v}{2}, \,\,
c = \frac{u-w}{d}.
$$
Condition (\ref{cond3}) is then satisfied if and only if $|a+c+bd|$, $|a+c|$ and $|a-c|$ are all $\le 1$.
For $s$ and $\hat s$ we have to insert the values of the optimal PPT state $\sigma$, obtained at the end
of the previous section.

After a tedious calculation, we get 6 conditions
an additive state has to satisfy for each of the three regions $A$, $B$ and $C$ of Figure 1.
Fortunately, only two of this total of 18 conditions
can be violated by expectation values belonging to normalized
positive states.
In the $A$ region all states are additive,
in region $B$ we must have $f\geq -2/d$,
and in region $C$ the condition is $\hat f \leq 3-4/d + (d-1)f$.
These conditions give us the border between the additive and non
additive areas.
\QED

\begin{figure}
\begin{center}
\epsfxsize=7.5cm
\epsffile{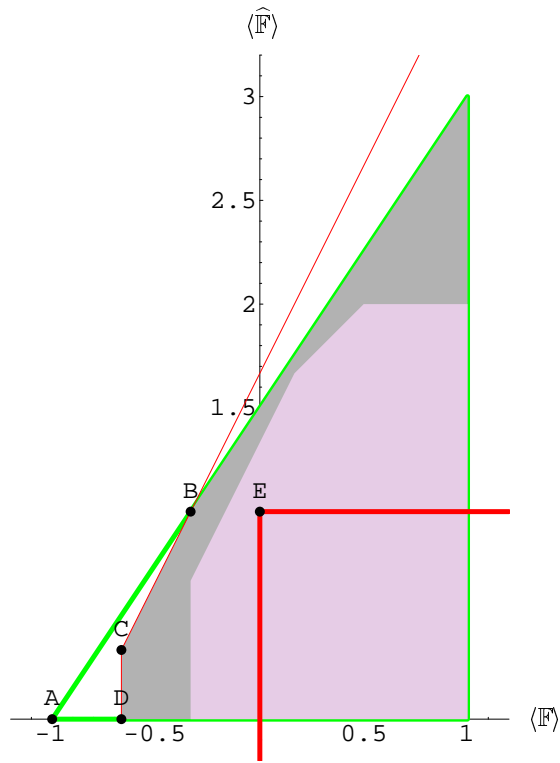}
\caption{ Additive areas for OO-invariant states (case $d=3$).
The state space has been subdivided in three regions. According
to Rains' Lemma, the states in
the light-grey region are strongly additive and those
in the dark-grey region are weakly additive. The region of additivity
is delineated by the line segments BC and CD. The points A, B, C, D and E
are defined in the text.
}
\label{fig2}
\end{center}
\end{figure}

The additive area for OO-states is plotted in dark grey in Figure
\ref{fig2} for the dimension $d=3$. States in the light grey area
fulfill the condition of strong additivity.

For later use, we have marked some points in the state space that
will become important in the further calculation of the AREEP. The
two additivity conditions of the Lemma correspond to the boundary
line segments CD and BC, respectively.
\begin{center}
\begin{tabular}{ccc}
  \hline
  Point & $\big< \F\big>$ & $\big< \flipt\big>$ \\ \hline
  A & $-1$ & $0$ \\
  B & $\frac{d-4}{d}$ & $d-2$ \\
  C & $\frac{-2}{d}$ & $\frac{d-2}{d}$ \\
  D & $\frac{-2}{d}$ & 0 \\
  E & $0$ & $1$ \\
  X & $\frac{4-6d+d^2}{d(d+2)-4}$ & $\frac{d^2(d-2)}{d(d+2)-4}$ \\
  Y & $\frac{-d^2}{d(d+2)-4}$ & $\frac{d(d-2)}{d(d+2)-4}$ \\
  \hline
\end{tabular}
\end{center}

\subsection{Rains upper bound}
In the second step we want to calculate the Rains bound (\ref{rns})
on the OO-invariant state space.
All OO-invariant states satisfy $|\sigma^{T_2}|^{T_2}\geq 0$, and we therefore restrict
the optimization to OO-invariant states $\sigma$. Since we want to use the Rains
bound as an upper bound, we need not to know that our so
restricted $\sigma$ is really the optimal one. But due to the
high symmetry of the OO-states it can easily be shown that
the optimum over all possible states $\sigma$ is
attained on OO-states anyway.

For additive states we have noted already that $E_R(\rho) =
E_R^\infty(\rho) = R(\rho)$ so that calculating the REEP directly
gives the Rains bound. To calculate the Rains bound in the
non-additive region ABCD, we have to perform the minimization
explicitly. Let the states $\rho$ and $\sigma$ be determined by
their expectation values $f, \hat f$ and $s, \hat s$,
respectively. Using the formula for the relative entropy of $\rho$
(\ref{relent1}) w.r.t.\ the optimal $\sigma$ for the REEP (see the
Table in Section \ref{sec_OO_sym}) yields the Rains Bound for
additive states.

For non-additive states we have to include the negativity of $\sigma$, given by (\ref{negativity}):
$$
\tr |\sigma^{T_2}| = \frac{|s|}{d} + \frac{|1-\hat s|}{2} +
\frac{d+d\hat s-2s}{2d}.
$$
As we will only use the above formulae for $\rho$ in the non-additive region ABCD, it is immediately
clear from Figure 2 that the optimal $\sigma$ will have negative $s$. We can, therefore, simplify the formula for the
negativity to
\beas
\tr |\sigma^{T_2}| &=& \frac{-s}{d} + \frac{|1-\hat s|}{2} + \frac{d+d\hat s-2s}{2d} \\
&=& \max(1,\hat s) - \frac{2s}{d}.
\eeas
Because of the `max' function appearing in this formula, we have to consider
two cases for $\sigma$ and, in the end, choose the solution that gives the smallest value for the Rains bound.

Consider first the case $\hat s>1$; then the negativity equals $\hat s-2s/d$ and we have to minimize
\beas
&& \log\frac{d\hat s-2 s}{d}
 + \frac{1}{2d}\Big[ 2 \hat f \log\frac{\hat f}{\hat s}
 + (d-d f)\log\frac{f-1}{s-1} \\
&& + (d+d f-2 \hat f) \log\frac{d+d f-2 \hat f}{d+d s-2 \hat s} \Big]
\eeas
over $s$ and $\hat s$.
This function has a single stationary point given by
\beas
s &=& \frac{d^2-d\hat f-2}{(d^2-2)f-d\hat f} \\
\hat s &=&\frac{-2\hat f}{(d^2-2)f-d\hat f}.
\eeas
However, the minimum we are looking for is a constrained one: the parameters $s$ and $\hat s$ must be
expectation values of positive $\sigma$.
On inspection, the positivity conditions are never satisfied in the stationary point for any choice of
$f,\hat f$ corresponding to a positive $\rho$.
Therefore, the stationary point is outside the feasible set (the state triangle) and
the constrained minimum will be found on the boundary of the feasible set.
This mere fact already rules out the present case $\hat s>1$, because
we know that the optimal $\sigma$ must be closer to the set of PPT states than $\rho$ itself, in the sense that
$\sigma$ should have lower negativity than $\rho$.
Indeed, setting $\sigma=\rho$ (which is certainly not optimal) in the
Rains bound yields a lower value than one would get for any $\sigma$ with a larger negativity than $\rho$.

We can, therefore, restrict ourselves to the case $\hat s \leq 1$.
As the negativity is then $1-2s/d$, the function to be minimized is
\bea
&&\log\frac{d-2 s}{d}
 + \frac{1}{2d}\Big[ 2 \hat f \log\frac{\hat f}{\hat s}
 + (d-d f)\log\frac{f-1}{s-1} \nonumber \\
&& + (d+d f-2 \hat f) \log\frac{d+d f-2 \hat f}{d+d s-2 \hat s} \Big].
\label{megaformel}
\eea
The stationary point is
\bea
s&=&\frac{2+d f}{d+2 f}\label{qq1} \\
\hat s&=&\frac{(2+d)\hat f}{d+2 f}.\label{qq2}
\eea
Again, $s$ and $\hat s$ must be expectation values of positive
$\sigma$ and we must have that $\hat s \leq 1$.
It turns out that the positivity conditions are always fulfilled. The condition $\hat s\leq 1$, on the
other hand, is only satisfied for states $\rho$ on or below the line going through points C and Y.
Therefore, the stationary point is the constrained minimum only
for states $\rho$ in the quadrangle AYCD.
This leads to the solution for AYCD:
\bea
R_{AYCD}(\rho)&=& \frac{1}{2}\big( (1+f)\log(d-2)-2 \log d \nonumber \\
&& -(f-1)\log(d+2)\big)\label{sol1},
\eea
which now only depends on the Flip expectation value $f$ and is an affine function of $f$.

For states $\rho$ in the remaining triangle CYB, the stationary point is outside the feasible set,
so that the constrained minimum will lie
on the line $\hat s=1$. Minimization of (\ref{megaformel}) over $s$, while fixing $\hat s=1$,
yields a quite cumbersome looking formula.
For later use, however, we will only need to know the resulting Rains bound on the line segment YB.
The solution consists of two cases, corresponding to either solution of a quadratic equation.
The end result is that, for the states on the segment YX, the Rains bound is given by
\bea
R_{YX}(\rho)&=& \frac{1+f}{2}\log d(1+f) \nonumber \\
&& +\frac{1-f}{2}\log\frac{d(1-f)}{d-1} \nonumber \\
&& +\log\frac{d(d+2)-4}{d^2}-\log 2.
\label{opt3}
\eea
For the states on the segment XB the bound is given by
\beq
\label{opt4}
R_{XB}(\rho)=\frac{1+f}{2} \log(d-2)+\frac{f-1}{2}\log \frac{d}{4}.
\eeq

\begin{figure}
\begin{center}
\epsfxsize=7.5cm
\epsffile{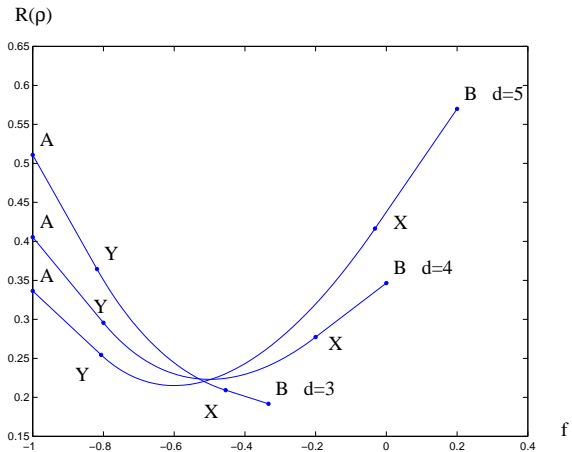}
\caption{ Rains bound on the line segment AB (see Figure 2) in terms of the parameter $f$,
for three different values of $d=3,4,5$. The bound consists here of
a linear part (segment AY, eq.\ \protect\ref{sol1}),
a curvilinear part (segment YX, eq.\ \protect\ref{opt3})
and again a linear part (segment XB, eq.\ \protect\ref{opt4}).
}
\label{fig22}
\end{center}
\end{figure}
Figure \ref{fig22} shows the Rains bound along the line segment AB, for several dimensions $d=3,4,5 $.
\subsection{Minimal convex extension}
\begin{figure}
\begin{center}
\epsfxsize=7.5cm
\epsffile{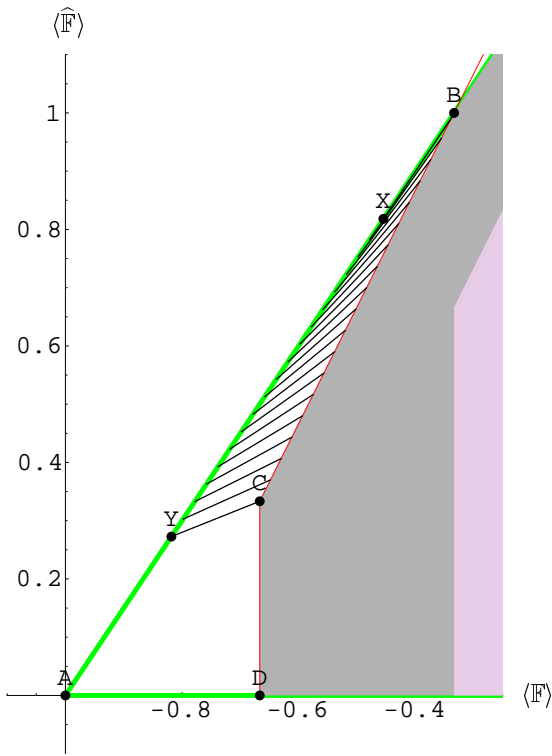}
\caption{A close-up of the non-additive OO-invariant states in
Figure 2, for the purpose of calculating the minimal convex
extension to the AREEP. In region AYCD, the minimal convex
extension only depends, affinely, on $f$ (eq.\ \protect\ref{ub1}).
In region BCY, the minimal convex extension is affine along the
lines depicted here (given by eq.\ \protect\ref{les}). }
\label{fig4}
\end{center}
\end{figure}

In this third and final step we calculate the minimal convex
extension of the additive area. This will turn out to be more
complicated than in the Werner states example. We will look at
straight lines, each connecting one point on the additivity border
with one, well-chosen point on the line segment AB.

The simplest case is the part of the additivity border consisting
of the line segment CD, because this line lies completely in the
`Werner' region, region $B$ in Figure 1, where, according to
(\ref{sol1}), the REEP only depends on the Flip expectation value
$f$. So, here, the two-dimensional problem is reduced to a
one-dimensional one. The REEP in the Werner triangle is given by
$$
E_R(f)=\log 2+\frac{(1+f)}{2}\log{\frac{1+f}{2}}+\frac{(1-f)}{2}\log{\frac{1-f}{2}}.
$$
As lower bound for the AREEP we get
\bea E_R^{\infty}(f,\hat f)
&\geq& E_R(-2/d)+(f+2/d)\frac{\partial E_R(f)}{\partial f}\big|_{f=-2/d} \nonumber \\
&=& \frac{1}{2}(1+f)\log\frac{d-2}{d+2}+\log\frac{2+d}{d}\label{ub1},
\eea
which happens to be identical to the Rains bound (\ref{sol1})
in the whole region AYCD. So the upper and lower bound equal
each other within this region and, hence, $E_R^\infty$ is equal to the Rains bound in AYCD.

The situation for the remaining triangle YCB is somewhat more complicated. To
calculate $E_R^\infty$ we consider a set of straight lines
connecting points on the line segment BC with points on the segment
XY and given by
\beq
\label{les}
\hat f= -p f+\frac{p(d^2-2+(d-2)p)}{2+d(p-2)-2p}.
\eeq
These lines are parameterized by $p$, which runs from $\frac{-2}{d+2}$ to $\frac{-d}{2}$.
Recall that the line XY is given by $\hat f=(1+f)d/2$ and BC by
$\hat f=3-4/d+(d-1)f$.

On the line segment XY, the Rains bound is given by (\ref{opt3}). On the segment BC, and in fact to the right
of it as well, the Rains bound is equal to $E_R=E_R^\infty$ and is given by
(\ref{relent1}) with $s=0$ and $\hat s=1$ (region $C$ of Figure 1). Moreover, this formula
holds for all points on the lines (\ref{les}) within the additivity region, allowing for the calculation
of the derivative of the Rains bound along the lines (\ref{les}). Doing this in the points on the additivity border
BC yields the result that, for every line (\ref{les}), the tangent to the Rains bound at the start point
(on segment BC) touches the Rains bound again at the end point (segment XY). By convexity
of $E_R^\infty$ and of the Rains bound, and by the fact that the Rains bound is an upper bound on $E_R^\infty$
and the tangent a lower bound, it follows that both $E_R^\infty$ and the Rains bound must coincide with this tangent
and, hence, be affine along each of the lines (\ref{les}). We conclude that $E_R^\infty$ is equal
to the Rains bound also in the remaining region YCB.
\subsection{Summary of results}
\begin{figure}
\begin{center}
\epsfxsize=7.5cm
\epsffile{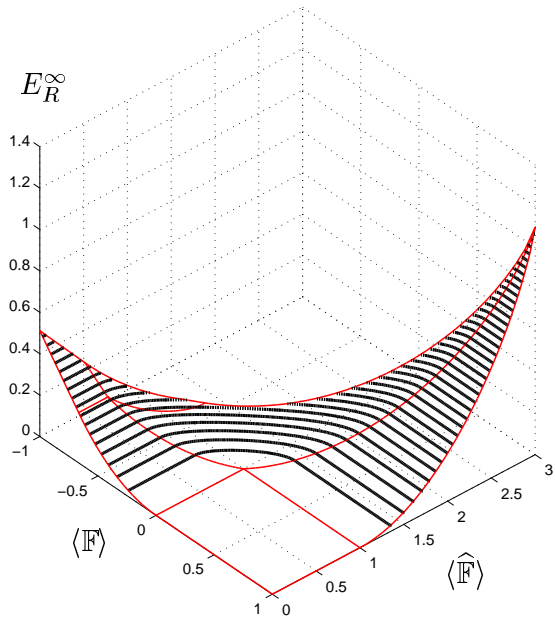}
\caption{Contour plot of the AREEP $E_R^\infty$ for the OO-invariant states,
parameterized by $f$ and $\hat f$ ($d=3$). Superimposed on this
plot are the lines separating the different regions defined in the text
(regions $A$, $B$ and $C$, the PPT set, the set of additive states and the regions AYCD and CYB).
}
\label{fig3d}
\end{center}
\end{figure}

We finalise the calculation of $E_R^\infty$ on the OO-invariant
states by summarising all the results obtained for the different
regions in the following table. Figure \ref{fig3d} shows a contour plot of $E_R^\infty$
for the case $d=3$.
\begin{center}
\begin{tabular}{ll}
  \hline
  Region & $E_R^\infty$ \\ \hline
  PPT & 0 \\
  $A$ & $E_R$, (\ref{relent1}), with $s=\frac{1+(d-1)f-\hat f}{d-\hat f}$ and $\hat s=1$ \\
  $B\setminus\text{AYCD}$ & $E_R$, (\ref{relent1}), with $s=0$ and $\hat s=\frac{\hat f}{1+f}$ \\
  $C\setminus\text{CYB}$ & $E_R$, (\ref{relent1}), with $s=0$ and $\hat s=1$ \\
  AYCD & eq.\ \ref{sol1} \\
  CYB & affine along lines (\ref{les}) between YX and BC\\
  YX & eq.\ \ref{opt3} \\
  \hline
\end{tabular}
\end{center}
Furthermore, the Rains bound is equal to $E_R^\infty$ in any of
these regions.

\section{Conclusion} \label{sec_disc}
In this paper, we have considered the calculation of the
AREEP $E_R^\infty$ for the class of OO-invariant states,
generalizing the results of \cite{ka}, which dealt only with the
class of Werner states. This has been achieved using four
basic ingredients: properties of the REEP $E_R$, properties
of the Rains bound $R$ (\ref{REntropy}), and a deep connection
between these two quantities $E_R^\infty$ and $R$. The final
cornerstone of the calculation is the symmetry inherent in the
OO-invariant states \cite{vollbrecht1}.

The relevant properties of the REEP are that it is an
additive entanglement measure in a large region of state space
\cite{Rains_Lemma} and that the AREEP is convex
everywhere \cite{rudolph}. This convexity allows us to use the
``minimal convex extension'' construction as a lower bound.

We have shown here that the Rains bound is also convex and
continuous, and that the calculation of it can be reformulated as
a convex optimization problem, which implies, by the way, that this
problem can be solved efficiently and does not suffer from
multiple local optima.

We have also made explicit the
techniques that were already employed in \cite{ka} implicitly, resulting in Lemma
\ref{RP}. This Lemma shows that there is a deep connection between
the AREEP and the Rains bound and seems to suggest that both regularise to the same
quantity \cite{rains_private}. Unfortunately, in its current
form, the Lemma is weakened by the additional requirement on the
states $\sigma$, over which the Rains bound is minimized, that
the quantity $|\sigma^{T_2}|^{T_2}$ should be positive. We have
coined the term binegative states for those states that violate
this requirement and we have made some initial investigations into
the question of their existence. Specifically, we showed that for
the case of OO-invariant states, $\sigma$ is not binegative, so
that the Lemma can be used here at full strength.
If it turned out that the extra requirement can always
be removed, in one way or another, then the Lemma could directly be used to prove
Rains' suggestion that $E_R^\infty = R^\infty$.

For the time being, we have been
able to show that at least for $E_R$-additive states $\rho$, the Rains
bound and the REEP are equal (and, of course, also equal to their
regularised versions).

Using these results, we have calculated the AREEP for OO-invariant states and it followed as a by-product
of the calculation that the Rains bound is identical to $E_R^\infty$ for
the OO-invariant states.

This last result could be taken as a hint that
the Rains bound might be additive everywhere, in contrast to $E_R$.
If this were true, then this would imply that the AREEP is
precisely equal to the (non-regularised) Rains bound and, furthermore, that it
can be calculated efficiently.

This work has been supported by
project GOA-Mefisto-666 (Belgium),
the EPSRC (UK),
the European Union project EQUIP,
the Deutsche Forschungsgemeinschaft (Germany)
and the European Science Foundation.
KA wishes to thank J. Eisert, M. Plenio, F. Verstraete, J. Dehaene and E. Rains for
fruitful discussions and remarks.

\end{document}